# Unmet Needs and Opportunities for Mobile Translation AI

**Daniel J. Liebling, Michal Lahav**
Google
Seattle, WA
dliebling,mlahav@google.com

**Abigail Evans**
Northeastern University
Seattle, WA
ab.evans@northeastern.edu

**Aaron Donsbach, Jess Holbrook, Boris Smus, Lindsey Boran**
Google
Seattle, WA
donsbach,jessh,smus,lboran@google.com

## ABSTRACT
Translation apps and devices are often presented in the context of providing assistance while traveling abroad. However, the spectrum of needs for cross-language communication is much wider. To investigate these needs, we conducted three studies with populations spanning socioeconomic status and geographic regions: (1) United States-based travelers, (2) migrant workers in India, and (3) immigrant populations in the United States. We compare frequent travelers' perception and actual translation needs with those of the two migrant communities. The latter two, with low language proficiency, have the greatest translation needs to navigate their daily lives. However, current mobile translation apps do not meet these needs. Our findings provide new insights on the usage practices and limitations of mobile translation tools. Finally, we propose design implications to help apps better serve these unmet needs.

## INTRODUCTION
Machine translation research began in the mid-20$^{th}$ century. Influenced by global conflicts and desire to understand content from foreign nations, multilingual aligned corpora facilitated development of translation systems with reasonable quality [21]. With the growth of the World Wide Web, early sites like AltaVista Babelfish brought machine translation to wider audiences. Most modern translation websites still retain the "two text box" interface introduced by SYSTRAN. As smartphone use grew, translation apps appeared with additional affordances including speech recognition and augmented reality [14].

Access to translation, then, appears to be universally available. However, little public research exists that details people's translation needs and examines how translation apps meet those needs. For what populations, in which scenarios, and what content is translated? How do translation apps succeed or fail at these tasks? This paper sheds new light on these questions. We detail the language needs of, and use of translation technology by, three different communities: English-speaking travelers based in the United States, Hindi-speaking intra-national migrant workers who live in Tamil-speaking Chennai, India, and immigrants with low English proficiency who live in Seattle, United States. The results of three related studies are integrated to accomplish two aims: (1) Investigate perceived and actual language needs in three different communities. (2) Detail the success and failure of translation apps to meet these needs.

In general, the Chennai migrants and US-based immigrants interviewed lacked effective communication tools and this had severe social, emotional and financial impact on their lives as well as on their sense of personal agency. For the US-based immigrant community, Major translation apps were frequently used, but were only useful for short, transactional communication. Their limitations included deficiencies in supporting long-form conversation, lack of vocation-specific vocabulary, significant translation errors, inadequate support for pronunciation and dialect differences, and lack of affordances tailored to urgent situations.

## RELATED WORK
*Immigrants' Use of Mobile Technology*
Prior work shows that some migrants utilize mobile technologies to support their unmet language needs [10, 32, 33]. However, studies of migrants' use of mobile and other technologies is limited because understanding of mobile language learning tools has focused predominantly on foreign language learners at the post-secondary level [26, 46] rather than those who must communicate in a non-native language to survive. Literature in the Mobile-assisted Language Learning (MALL) space identifies the need for authentic synchronous communication [10, 27, 32]. A variety of mobile translation services and dictionaries have aimed to address the communication needs of immigrants, namely lacking vocabulary knowledge, lacking pronunciation capability [10, 11], and the inability to communicate [13]. However, Burston notes lack of proliferation outside educational settings [7]. Adoption of translation technology tends to be a mentioned only in passing in most information and communication technologies for development (ICT4D) literature. Similar patterns appear similar across populations, with language learning and translation apps noted most often. Both Korean [44] and Latina [1] immigrants in the United States used online and/or mobile translation and dictionary apps. Use of these apps can become part of people's routines; Lingel reports an individual regularly using Google Translate prior to purchasing groceries [28]. We aim to do a deep dive into understanding the perceived and actual needs of two immigrant communities and investigate the successes and failures of translation apps to meet these needs.

*Translation App Modalities*
Modern translation apps provide multiple modalities for input and output including text entry, speech recognition, and text-to-speech. In their meta-review, Clark et al. [9] note that evaluation of speech interfaces in real-world contexts is limited. Similarly, *in situ* evaluation of speech-to-speech translation systems with natural conversations is also understudied. Hara and Iqbal [20] performed a laboratory study of speech-to-speech translation with Microsoft's Skype Translator, but used actors. They found that participants preferred text transcripts to text-to-speech output because the text provided a temporary context useful for overcoming errors. The use of text provides additional confidence-checking; people not only wanted to know the gist of source text in translation, but also wanted a word-by-word translation so that they could independently verify the accuracy [1]. Text transcripts can also aid conversation dynamics; live transcription can improve perceived conversation quality and comfort [17, 34]. Although most popular translation apps include copy and paste of translated text as a feature, users still have difficulty with copy and paste functionality on mobile devices [23] for reasons including discoverability [15] and accessibility.

*Privacy Among Immigrant Populations*
Guberek et al. found that undocumented immigrants articulate complex views of technology and privacy [19]. The degree of personal privacy of mobile devices varies around the world. For example, Sambasivan et al report monitoring of women's phones by other family members in India [37]. Sensitivity to input and output modalities also varies; Fukumoto notes that voice input is rarely used in Japan due to limited access to private spaces and cultural bias against extra-conversational speech in public [16].

Our work presents the results of three related studies which accomplish two aims. We investigate perceived and actual language needs in three different communities. We detail the success and failure of translation apps to meet these needs. The remainder of this paper is organized as follows: we first present survey results from frequent travelers in the United States. We then discuss findings from a diary study and semi-structured interviews with intra-national migrants in Chennai, India. Third, we detail semi-structured interviews with international immigrants to the United States. Finally, we contrast the needs of travelers with those of the two migrant populations, summarizing unmet needs and proposing design implications.

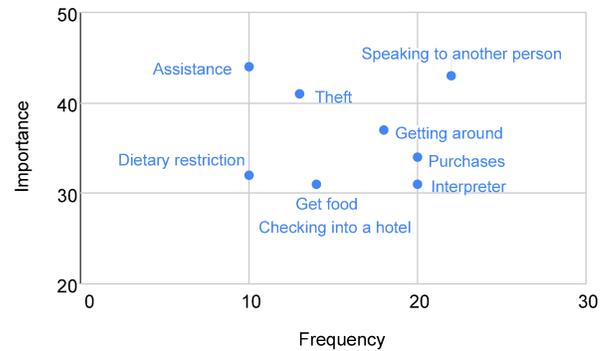

Figure 1. Perceived importance (% of responses rating a scenario moderately important or higher) vs. frequency (% of responses encountering a scenario half the time or more).

## FREQUENT TRAVELERS FROM THE UNITED STATES

Our first study investigated the perception and importance of translation scenarios when traveling from the United States to another nation where English is not an official language. In the United States in 2018, approximately 42% of the population held passports [48] and 19.2% of air departures landed in Europe [49]. Hence, travel-related translation needs are easy to conceive. Scenarios such as asking directions and ordering food in restaurants are commonly advertised use cases for a variety of recent consumer speech-to-speech translation devices. To disentangle the perceived need from the actual need, we surveyed a representative group of travelers.

### Methods
Because we were interested in mobile use of translation while traveling, we restricted eligibility to respondents who have a smartphone and who traveled somewhere where they did not speak or understand the local language within the past 2 years. 3,105 US respondents (261 to 364 unique participants per survey) ages 18 and older completed Triptech–style online surveys [39] in exchange for access to premium content on a network of sites.[1]

Each respondent was asked about the frequency and perceived importance of addressing one of the nine situations listed in Table 1 by answering two questions:

*Importance.* How important is it to address the following problem when traveling and you do not speak the local language? [Not at all, Somewhat, Moderately, Very, Extremely]

*Frequency.* How often do you encounter the following problem when traveling and you do not speak the local language? [Never, Sometimes, Half the time, Most of the time, Always]

Figure 1 shows a chart of perceived importance and frequency for responses rating each scenario important (moderately important or higher) and frequent (half the time or more). When asked about perceived importance of communicating in the local language, 54% of respondents selected that it is at least moderately important, with 27% indicating that it is very or extremely important. Over 40% of these responses rated emergency situations (medical and police) important, but less than 15% of responses indicated these occurred frequently. The most important and most frequent scenario was speaking informally with another person (43% important; 22% frequent). This is in contrast to needing a more formal interpreter which was seen as important in fewer responses (31%), but roughly the same in frequency (20%). Since 54% of responses indicated that communicating in the local language was important, there may be missing scenarios or framing effects in the survey that contribute to the 11% gap.

---
[1] https://www.google.com/surveys

| Scenario | Prompt |
|---|---|
| Speaking with people | I need to speak to someone who speaks another language than I do. |
| Getting around | I need to ask for directions but I don't speak the local language. |
| Purchases | I need to buy something but I don't speak the local language. |
| Checking into a hotel | I need to check-in to my hotel but I don't speak the local language. |
| Get food | I need to buy food but I don't speak the local language. |
| Dietary restrictions | I have a food allergy or preference I need to tell someone about but I don't speak the local language. |
| Assistance | I need medical assistance but I don't speak the local language. |
| Theft | I need help from the police but I don't speak the local language. |
| Interpreter | I need a language interpreter or guide to help me communicate. |

**Table 1. Scenarios rated by respondents on dimensions of importance and frequency.**

The survey data indicate that, for travelers from the United States, travel-related scenarios are generally seen as important, but not frequently encountered. Independent of frequency, most of these scenarios are transactional in nature. They require only direct translations of highly scripted interactions [41]. Dictionary lookup and slot-filling strategies will satisfy these communication needs in many cases. Fully phrase-based or neural machine translation is not required. Many popular translation apps already support phrase tables. For example, Microsoft's mobile translation app allows users to pick a source-language phrase, fill empty slots (named entity, quantity, etc.), and output the target translation. The Google Translate app also has a phrasebook feature, and frequently used translations are edited and verified by an online community. The scenario that was rated as both most important and most frequent (upper-right quadrant of Figure 1) was the scenario that translation apps support least well: conversation. State of the art multimodal dialog management systems require extensive computation power [5] and are not yet available on mobile devices.

In the next two studies, we reveal how important these scenarios are for two different sets of migrant populations. Facilitating cross-language conversations for these individuals is fundamental to gain access to services and reduce isolation.

**MIGRANTS IN INDIA**
Transnational migrants often encounter language barriers, but intra-national migrants also face these challenges. To learn more about cross-language communication needs of intra-national migrants and how they use translation technology, we executed in-home, semi-structured interviews with migrants in Chennai, India. Although Hindi is widely spoken and/or understood in India, Tamil is the official language of Chennai, employed by approximately 69 million native speakers and 8 million second-language (L2) speakers [12]. Chennai and its surroundings are home to over 7 million people, with an estimated 1 million migrant workers [35].

**Phase 1: Revisited Structured Interviews**
*Methods*
We employed a local research firm to recruit Chennai-residing migrant workers who did not speak Tamil but spoke Hindi fluently. That is, they spoke a minority language and not the majority language of their own country. The facilitator recruited participants in a public location and administered the questions orally to accommodate participants with low literacy.

16 participants (12 male) aged 20–42 years completed structured interviews, for which they received incentives. Despite our best efforts, the participant pool was biased to male participants primarily due to the following cultural factors: Participants were interviewed in public spaces that typically had many more males than females. Women who were on their own when approached were reluctant to be interviewed without their husband present. Sometimes, when a woman's husband was asked for permission, he declined. To reach more female participants, we used snowball sampling, e.g. one participant was recruited because she was related to another female participant. We gender-balanced the sample for our in-home interviews by following up with all of the women but not all of the men. Table 2 summarizes participant demographics. Respondents detailed the locations where they encountered language barriers, the frequency of those encounters, and any strategies that they developed to overcome language barriers. The facilitator asked about and transcribed respondents' descriptions of recent language difficulties.

Each respondent was sampled every two days for a total of three sessions. After reviewing the Phase 1 study results, we performed semi-structured interviews in the participants' homes, detailed later in this section.

*Findings*
Phase 1 revealed two major types of language barriers: transactional (commerce and work) and survival (access to education and healthcare). Participants agreed that lack of proficiency impacted these areas. Eight strongly agreed that finding work was adversely affected; 12 strongly agreed that shopping or healthcare was. In transactional scenarios, respondents noted difficulty in arranging relatively simple affairs:

> "I went to a mixer repair shop. I could not explain to the shopkeeper that I wanted to change the mixer knob. But the shopkeeper did not understand my language. I even tried sign language but could not explain to him. Finally I left the shop and could not repair my mixer." —P2

| | Interview? | Gender | Age | Place of origin |
|---|---|---|---|---|
| P1 | ✓ | F | 30 | Bihar |
| P2 | ✓ | F | 25 | Bihar |
| P3 | | M | 32 | West Bengal |
| P4 | ✓ | M | 23 | West Bengal |
| P5 | ✓ | M | 21 | Odisha |
| P6 | ✓ | F | 33 | Bihar |
| P7 | | M | 30 | Bihar |
| P8 | ✓ | F | 35 | Bihar |
| P9 | | M | 23 | Assam |
| P10 | | M | 42 | West Bengal |
| P11 | | M | 21 | Odisha |
| P12 | | M | 24 | Bihar |
| P13 | ✓ | M | 20 | West Bengal |
| P14 | | M | 23 | Assam |
| P15 | | *information not provided* | | |
| P16 | ✓ | M | 28 | Rajasthan |

Table 2. Chennai, India participant demographics.

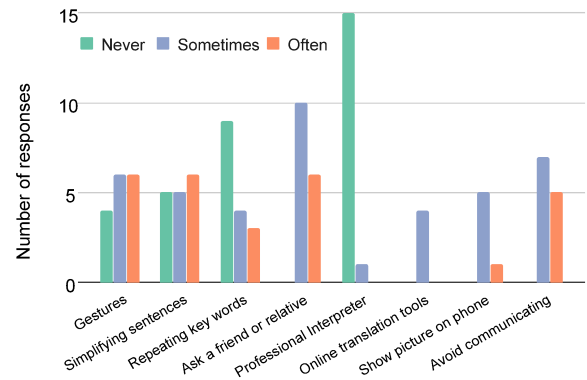

Figure 2. Phase 1 participants' strategies for overcoming language barriers.

Using a known or even spontaneously-recruited multilingual contact to interpret was the most common strategy to overcome language barriers and used by all respondents. This was done in-person or by phone. Figure 2 shows the frequencies of other strategies.

> "The customer ordered some dish which was not available in the restaurant, but [was] am not able to inform this to the customer, he could not understand what I was saying since I was speaking in Hindi, finally I asked help from my ... friend who knows Tamil to inform the customer." —P9

> "I was ... on the phone. I wanted to change ... where he delivers the newspaper but he could not understand what I was saying. First I spoke to him in Hindi and then in English he could not understand. I tried a lot. Finally the newspaper man gave the phone to someone else who could understand English. I told him about the change of address and he then told the newspaper man." —P8

Having to rely on strangers to interpret can cause a perceived or real loss of status. Some participants noted a reluctance to speak in public because not speaking Tamil would identify them as a lower-status individual. The use of strangers also creates a loss of privacy:

> "I went to the hospital yesterday. The [receptionist] was not able to understand my problem... She doesn't know Hindi. Finally, another patient helped me... She translated my problem to the [receptionist]." —P5

### Phase 2: Semi-Structured Interviews

*Methods*

After reviewing the Phase 1 studies, we chose 8 participants for in-home semi-structured interviews. We selected the interview subjects to balance across occupation (e.g. students, homemakers, and tradespeople), gender, and age. 4 male and 4 female participants sat for single interview sessions conducted in their residence (see Table 2). The average age of those interviewed was 27 years, close to the average age of migrant workers in Chennai, 29 years [22]. Each session lasted approximately 60 minutes. A single English-speaking monolingual researcher communicated with each participant through a Hindi-English bilingual interpreter. A Hindi-speaking Indian researcher and one or more English-speaking American staff observed the interviews and transcribed the English turns of the conversation.

The interpreter briefed each participant on the interview process and gained consent to use the interviews for research purposes. All participants were compensated in cash by a third party. Approximately 45 minutes of each session was inquiry and response to a set of predetermined questions, informed by Phase 1 studies results, and designed to elicit additional context. In particular, we sought to contextualize communication difficulties within the environments that the participants lived: their homes, their families, and their workplaces.

*Findings*

The interviewees elaborated on the Phase 1 studies and consistently expressed difficulty conducting routines of daily life such as purchasing items, ordering food, and resolving simple conflicts. Lack of language proficiency often manifested as anxiety around conversation. P6 said: "I don't know Tamil. I always have this fear inside while talking to people."

*Learning another language.* Many participants had lived in Chennai for more than three years, with one as many as 15 years. Nonetheless, almost all viewed their residence in Chennai as temporary. This seemed to influence their views on learning Tamil. Some did not realize that learning Tamil would be so challenging. P5 stated that when moving to Chennai, they "did not think of the language issue and did not know it would be so difficult." At least three participants attempted to learn some Tamil, but all cited lack of time as a barrier.

Many mentioned that if they were to learn a second language, it would be English. Use of English carries cachet as it is associated with the upper castes, so acquiring English is a form of upward mobility [2].

> "If I learn English I'll be more respected and if I go to town, many educated [people] speak more English than Hindi, so I'd be able to converse with them." —P16

*Commerce.* Almost all participants described concern about being overcharged and missing out on discounts. Unable to negotiate prices or assert themselves, they regularly experience monetary losses:

> "When I go for dinner [I] cannot interact with them properly. I use gestures. I ask what's the cost, they reply in Tamil. I just hand them a ₹100 note and they hand change back to me." —P8

> "...the ticket was for ₹45 but since the conductor did not have change; he told us he will give it [back] before our stop comes. [We] both don't know the local language. When [we arrived at] our stop I told the conductor to give me the ₹5; I kept on telling him in Hindi that he has to return the money he just kept replying in Tamil. Since we could not understand we had to just get down without getting the change back." —P11

*Communication strategies*

As in the Phase 1 studies, the interviewees expressed a variety of strategies for dealing with language barriers. Children were frequently used as interpreters, which is consistent with other ethnographic studies [30]. Most families had children in government or private (but free) primary schools where they learned Tamil, English, and Hindi. Parents sometimes suspected that the children would translate incompletely especially in the context of parent-teacher conferences. Consistent with the Phase 1 studies, some parents simply avoided those scenarios because they felt like there was no possibility of communication.

Participants described calling a contact on their mobile phone to interpret a conversation. This sometimes involved handing their phone to a stranger. This is at odds with Karlson et al [24] who found that physical security and privacy concerns influenced the first party's willingness to hand their phone to a second party. However, the difference may be attributed to cultural practice or resolution of needs outweighing the risks. Guberek et al discuss such risk tradeoffs [19]. In one case, the individual's go-to interpreter was also their boss, which put them at risk for exploitation.

*Mobile technology use*

All participants used mobile phones but their type and use varied highly. Five interviewees had feature phones and three had smartphones. Of the latter, all participants mentioned media consumption in apps like YouTube and UC Browser. However, many individuals had low literacy skills. One mentioned that "everything [on the phone] is written in English ... I can't read it." Another noted that they used WhatsApp "only to call people or send videos" with family members, not to send text messages. One feature phone user was numerate, but did not know how to initiate a call on her phone; she only knew how to answer when her partner called her.

Despite difficulty communicating, many individuals did not aspire to learn Tamil. They often considered their residence in Chennai as temporary even though some had resided there for 10 years or more. Among those who attempted to learn Tamil, most did so through friends. For those with smartphones, most did not use apps due to the literacy barrier. However, a few made extensive use of apps for language learning. One person used Google Translate to communicate in Tamil by typing in English, listening to the Tamil translation through text-to-speech, then trying to reproduce the pronunciation. If that failed, he showed his phone with the text to the listener, who would correct his pronunciation. Another used the Shabdkosh dictionary app, which also includes pronunciation. However, they noted that the Tamil pronunciation "sounded like a robot" (due to use of a vocoder). At least two participants mentioned the UC Browser app, a mobile browser with *in situ* language-learning content including a "quiz mode" as well as audible pronunciations.

**IMMIGRANTS IN THE UNITED STATES**

More than 44.5 million immigrants[2] resided in the United States in 2017 [47]. Approximately 48 percent (21.2 million) of those immigrants ages 5 and older were Limited English Proficient (LEP) [6], suggesting a strong need for English language support and services. Government and other language resource programs exist to provide infrastructure for these language needs [11]. However, post-immigration difficulties indicate that these programs do not meet their daily needs [13]. Immigrants face many challenges [18] including acculturation and integration [4], employment [3], health [8, 29], and parenting [45].

To complement the Chennai study, we wanted to investigate a geographically different immigrant community and shed light on the what roll translation technology plays in the daily life of US-based LEP immigrants. Little is known what role mobile translation technology plays in the daily lives of immigrants. Our research questions were threefold: How do people with low English language proficiency navigate their daily challenges? What is the role of translation technology in navigating these challenges? How can we build solutions to address their needs? To investigate, we conducted semi-structured interviews in a major US city with 9 LEP immigrants.

**Methods**

*Participants*

We spent several months developing relationships with two local immigrant resource organizations and community center allies to assess how we could best recruit participants while protecting their privacy. Several allies then reached out to their communities for our study through their networks. Ultimately, our participants were 4 Spanish-speaking, 4 Mandarin-speaking, and 1 Russian-speaking LEP immigrants (4 women, 5 men). Despite our best efforts to seek participants with diversity in age, gender, and occupation, our sample (see Table 3) was dominated by participants in the 51–60's age bracket.

---

[2] Foreign-born residents

|    | Gender | Age group | Place of origin     |
|----|--------|-----------|---------------------|
| P1 | M      | 51-60     | Russian Federation  |
| P2 | F      | 51-60     | China               |
| P3 | F      | 51-60     | China               |
| P4 | M      | 51-60     | China               |
| P5 | M      | 41-50     | Vietnam             |
| P6 | M      | 24-30     | Columbia            |
| P7 | M      | 41-50     | Honduras            |
| P8 | F      | 60+       | Mexico              |
| P9 | F      | 51-60     | El Salvador         |

Table 3. United States participant demographics.

The Spanish-speaking participants emigrated from Mexico, Honduras, El Salvador and Columbia. Of the 4 Mandarin-speaking participants, 3 were from mainland China, while one was from Vietnam. The single Russian-speaking participant was from the Russian Federation. They are settled immigrants, who have been in the United States 0.5—18 years (median 8.2).

We focused on recruiting LEP immigrants to understand the most extreme use cases: The needs of individuals who struggle the most to understand the dominant language. All individuals were screened for LEP using the Interagency Language Roundtable (ILR) scale.[3] This consists of five levels of language proficiency; we screened for people with ILR levels of 0 and 1. Individuals at these levels are "unable to produce continuous discourse except with rehearsed material." The interviews were conducted in-person in spaces trusted by them (e.g., community center or library). Individuals received cash equivalent incentives through a third party. A researcher conducted the interviews through an accompanying consecutive interpreter. Prior to the start of the interview, each participant acknowledged an informed consent document provided in their native language and were allowed to decline while still retaining the incentive. Each interview lasted approximately 90 minutes.

*Data handling and coding*
We followed a strict security protocol in handling participant data. We kept consent forms with personally identifying information separate from the data collected in the study. Participants received a participant number to identify them, instead of their name. If participants agreed to video and photos taken, all files were stored in our institution's cloud storage, which is certified for sensitive identifiable human subjects data and has strict retention and deletion guidelines. We specifically refrained from asking participants about their context of departure/arrival and requested they not discuss any aspects concerning their legal status. All interviews were translated by one of two professional bilingual interpreters and notes were taken by the researcher. All potentially identifying information was redacted in the transcription process. Qualitative analysis was conducted on the redacted, English transcripts. Using thematic analysis, we explored the categories to identify and analyze prevalent themes.

---

[3] https://www.govtilr.org/Skills/ILRscale1.htm

**Findings**
We reviewed the transcripts and clustered participants' responses into the following taxonomy:

1. Social and emotional context
    (a) Stress, isolation and lack of personal agency
    (b) Social networks of support

2. Usage practices: Occupation
    (a) Occupational translation needs
    (b) Employment seeking
    (c) Professional downward mobility
    (d) Negative employer relations

3. Transactional needs
    (a) Commerce
    (b) Medical needs
    (c) High stress scenarios

Drawing from the interviews, we present concrete examples for each category in the taxonomy.

*Social and emotional context*
Overall, immigrants lacked effective communication tools and this had severe social, emotional and financial impact on their lives as well as on their sense of personal agency.

*Stress, isolation and lack of personal agency.* Despite their tenure in the United States, all participants exhibited separation acculturation [4], rejecting the dominant host culture in favor of preserving their culture of origin. Furthermore, they remained socially within their ethnic enclaves, limiting their integration. None reported having native English speaking friends. Participants reported feeling isolated, powerless, and having ongoing stress. P1 shared: "I can't speak to interesting people. I can't teach people, I can't pass my experience on to others. I can't affect society or economics. I can't express my opinion... I can't get a job... I feel powerless; I can't even defend my family with words." While P9 shared their feelings of isolation: "I feel like I'm in a desert. There are many people around me, but I am all alone." Participants attributed the lack of command of the English language as the root of these feelings of stress; P8 described this as "being *mute* all the time." This is in line with a corpus of literature describing the stresses commonly faced by immigrants and refugees [4].

*Social networks of support.* Participants' social networks were vital for day-to-day survival, but they served as a means for short-term coping, rather than long-term integration. They interacted socially with and acquired resources from those speaking their native language. That included renting housing from and/or with same-language speakers and going to hospitals that had fluent doctors or interpreter support. In line with the literature, all participants described having an informal assistant within their social network who would help with tasks such as going with them on shopping trips, helping them interpret legal documents, helping them read emails from the bank, billing statements, letters from children's school, etc. These ranged from family members, to roommates that would receive some incentive (e.g., gas money) for accompanying them on a shopping trip. The favored-friend strategy was particularly common with individuals for whom mobile translation apps did not meet their needs.

> "Have friends that have lived here a long time help fill out documents — in a bank with difficult questions — [I go] through friends. Have to figure out their schedule"
> —P3, Mandarin

> "When I came here when I had to apply for social security; I had to call my brother-in-law and he was the one who also helped me for applying for Green card and ID."
> —P4, Mandarin

However, even this resource network left them feeling frustrated and lacking personal agency. P6 described using their brother for translation assistance in an interaction with their landlord, where he felt lost, not understanding the nuances of the interaction:

> "My brother is our translator. Where we rent, he is the one who always talks to the owners. In those instances, I feel frustrated because I cannot say what I need or feel. I really don't know what my brother is saying or not saying. The water heater was broken. My brother was talking and the landlord was angry — I didn't know what part of what my brother was saying that made him angry."
> —P6, Spanish.

*Usage practices*

*Occupational translation needs.* Most participants experienced the bulk of their translation needs and challenges in the context of their occupation. The areas that highlighted the shortcomings of the translation technologies were (1) employment seeking, (2) decrease in their professional level within the labor market, or "professional downward mobility" and (3) navigating negative employer relations.

*Employment seeking.* Participants described the challenges of trying to find a job from employers that spoke their language, to try to bypass the need for the mobile translation tools in the interview process. One participant sought out an occupation that didn't involve in-person interviews, e.g. being a driver for a transportation app provider where they primarily used web based translation tools to navigate the on-line interview and registration processes. Our recruiting process biased our participants to those who specifically seek out community center resources, and thus a majority of our participants (5 of 9) described finding work through those employment resources. One participant described feeling fearful of the job interview and consequently bringing his English proficient son:

> "When I wanted to get a part time job, because I cannot understand English, I cannot do it. For [the] application, I could ask for someone else to help me, the most difficult part would be the interview. Earlier I applied for a job through the employment agency— for the interview— I brought my son for the interview. My son interpreted for me. Nothing came out of it, they didn't even give me an answer." —P5

*Professional downward mobility.* Consistent with previous research [8], many immigrants experience professional downward mobility. Nearly all participants reported to have significantly downgraded their professions since moving to the US, attributing the change to language constraints: Aeronautical engineer now doing maintenance and yard work; structural engineer, now a factory worker; mechanic now cleaning hotels. P1 described the frustration of starting from the beginning: "I had an engineering degree to do structural engineering, and I can't do any of that here. I worked on a factory that made fighter jets, here I can't even work for Boeing. When I moved here, I knew that I would be starting from scratch, from zero, like a boy." Participants attribute this struggle to their command of the English language. P7 "had several job opportunities, but ... refused to take them" because they "[weren't] able to communicate." Although previous researchers have examined occupational shifts, the role of translation technology in occupational changes has not been considered.

*Negative employer relations.* P9 described arriving at a house cleaning job and discovering hazardous conditions. She felt it was out of her comfort zone. She was able to use the mobile translation tool, but not enough to articulate and advocate for her boundaries. She expressed fear of abandoning a professional opportunity:

> "I managed to use the translator [app], but she was asking me to do things that weren't part of that work order... I decided to do most of the things she was requesting — to avoid the complaint — I decided to do what she was asking — I didn't want to get punished."

Participants voiced their lack of understanding around employer benefits. P3 demonstrated how this lack of understanding could derail them from the system: "When I was working at the hotel...they were going to get insurance and 401(k)[4] to carry over to my next job and to this day I don't understand."

*Transactional needs*

*Commerce.* Participants used a mix of tools when shopping: mobile translation tools, gestures, and recruited helpers. P5 told how she showed the translated text to a salesperson: "If I wanted to buy pants, I will input Chinese into the tool and then have it translated into English, show it to sales people, ask them where to find the pants." When asked what they felt they were missing out on by not speaking English, the top response was understanding the details of sales and discounts. P3 stated: "Sale, or discount, we receive these ads at home, however, because I don't understand it, I forget about it." They felt they didn't understand the parameters of the discounts and thus couldn't take advantage of them. In this context, participants felt they lacked agency to be savvy consumers and have an understanding of the rules of the world around them. P4 describes this disadvantage: "If there is a sale, on the day— if you know English you can buy things that are less expensive. These are situations where I need to know English— knowing would be an advantage."

*Medical needs.* Participants fell into one of two camps regarding their practices of seeking medical services. Two partici-pants said they avoided it; P7 stated: "I am never sick, so I don't know anything about that and I don't want to learn." Another felt intimidated and bewildered by the medical benefits system. P1 said they "don't go to doctor. We can't explain the sickness and won't understand what the doctor said. We still don't understand the whole [health insurance] system doesn't make sense... I think we were allowed into the states because we are healthy people." The majority of participants sought out clinics and hospitals with interpreter support. P2 noted "when I go to the hospital, or to see the dentist, the hospital has interpreters— no matter where I go I ask for interpreters." Another unmet need was phone call translation for making and scheduling follow-up appointments. Two participants mentioned taking public transport to their community resource center and waiting in lines, just to make these phone calls with an on-duty interpreter.

---
[4]Retirement savings program in the United States

*High stress scenarios.* Despite the ubiquitous usage of mobile translation devices, participants outlined circumstances where physically engaging with the phone to facilitate translation was inappropriate or not possible. In a tear-filled account, P8 described being assaulted and ultimately pepper-sprayed at work. Unable to communicate in English, she yelled for help, catching the attention of a fellow employee. Police were called to the scene and a police report was taken down from the only English speaker in the room: the alleged perpetrator. Later, her attorney interpreted the police report for her. To her devastation, the report was written erroneously, faulting her for pepper spraying the man. For another participant, the act of engaging with law enforcement was paralyzing. P6 recounted a story of being delayed for hours due to a minor traffic infraction:

> "Police stopped us and I didn't know what to do. Fifteen patrols came until one came that spoke Spanish. I was scared; I was not understanding what they were telling me and they were not understanding me either. You know how the law is here. You cannot move or do anything; I decided to stay still. I thought it was best to stay still."

**Use of Translation Technology**

The use of smartphone translation technology played an essential role in participants' lives. All but one participant used Google Translate on their phone device as their primary mode of communication with individuals who didn't speak their language of origin. The one participant that did not use Google Translate had no literacy in both English and language of origin, and thus was unable to engage with the app. P6 stated: "the truth [is that when] you arrive here, the only tool you have is the cell. I use Google Translate. I use the translator for everything— it's my right hand." In terms of the features used on Google Translate, eight participants used voice input feature, seven used camera/scan translation and text input features and four used 'conversation mode' feature. Mobile translation applications were ubiquitously used for short transactions such as asking for something at a restaurant or asking for directions. However, when it came to having longer, more nuanced conversations, participants felt the translation tools had limitations that severely impacted their life.

*Camera and Scan functions in translation apps.* Several mobile apps contain augmented reality translation features [14] that render translated text over the source text in a live camera feed. 7 of 9 immigrants used the camera translation functions to translate shopping scenarios (7 of 7) and document scanning like utility bills (5 of 7). These features afforded participants some narrow contextual translation: "Sometimes when I receive letters, rather than ask volunteers to look at it, I use the cell phone for that purpose." However, translations were often inaccurate. P3 noted that "the translation is not always grammatically correct, not very fluent or fluid, so I have to do some guess work in order to make sense out of it." The benefits seemed to outweigh the frustrations of untangling the meaning; P3 added that "the sentences may not make complete sense, but [they could] basically understand."

*Social acceptance of translation apps.* Participants spoke of the challenges and inhibitions using translation technologies. P1 described that asking someone to use translation technology to facilitate his own understanding was "disrespectful" and akin to "stealing their time." As in the Chennai study, handing the phone to a stranger seemed acceptable to most, in particular when a misunderstanding had gone on too long. P4 described, "when the person has said it several times and I couldn't understand, that is when I give the cell phone to that person, then they type it in, then I see the translation." More common, however was showing the translated text to a stranger. "I will just type it in, then it will translate and I will show it to that person." This was particularly true in scenarios when there is a contextual image, i.e., when people asked for directions. P3 accounts: "Use cell phone to type in the address and show it to people." Despite the benefits the mobile translation tools afford for short transactional communication, when one or more participants engage with the screen, they disengage from visual contact with each other. Participants' preference for text concurs with Hara and Iqbal [20] who found that users preferred text transcripts to Text-to-Speech audio because they help users overcome recognition or translation errors by providing contextual clues.

Participants emphasized that another barrier to more long-form conversation was the other speaker's acceptance of the device as a communication tool. P9 describes the elderly man she care-takes being impatient with her use of translation technology:

> "I have lost work— I have lost good clients because they don't have the patience to communicate with me. One of the clients told me that using translator— you work very well, but I need someone that I can communicate with."

*Translation Apps in Conversational Contexts.* Google Translate's *Conversation* feature, which allows for users to explicitly select their language and provide speech input which is then translated, similar to a chat interface, was used by four participants. The flow of the conversation, already stressed by latency of the recognizer and translation systems, becomes even more broken. P9 also spoke about the shortcomings of translation tech in the context of a romantic relationship:

"You feel like you are talking with a robot; you don't have the feeling of rapport. When you have love, it has a human flavor. You know how love is magic? When you say something and you have to wait and the other person responds, the magic is gone."

People voiced the desire for impromptu casual conversation without the barriers of a device. P6 explained, "of course, it would be a pleasure to be able to speak with someone on the bus. At the bus [stop], we just stare at each other because we cannot communicate."

### Limitations of translation technology

*Translation errors.* Most participants emphasized the frequency at which translation output was inaccurate, did not make sense or was simply incomplete. P3 said, "The translation is not always grammatically correct, not very fluent or fluid, so I have to do some guess work in order to make sense out of it." Sometimes these situations were merely frustrating, but in others the whole conversation could be derailed (translation is inappropriate or lewd) or even dangerous (directions or instructions are wrong, medical terms inaccurate). P9 worked as a housekeeper and describes the awkwardness of inaccurate translation: "Sometimes it gets bad words" (obscenities). "I touch and say it again, sometimes twice — it is frustrating, and I do it several times."

*Accent and dialect support.* The difficulties pronouncing English came up in many interviews. P7 noted that "English pronunciation is *very* hard. If the mobile translation tool is not trained on accented speech, the app will error." P5 confirmed this: "If your pronunciation is not accurate the translation will not be as accurate. In China, there are many different dialects. This happens with English as well, when you speak into the English, it doesn't come out as accurate. Sometimes the sentence is reversed; the translation is reversed."

*Vocation specific vocabulary.* One of the major shortcomings of mobile translation tools in a professional context was the vocation-specific vocabulary participants felt was often missing or incomplete. For example, one participant that was a construction inspector as his prior-to-immigrating occupation, but with lack of construction specific vocabulary, he felt unable to rely on mobile translation tools.

### Language learning and translate apps

Most participants described struggles and stress around the difficulties of language learning. P7 lamented "I can tell how behind I am in everything — there was a time I used to stay at home and cry a lot — I realize I don't know how to say everything." Almost all participants spoke of a pressure to accelerate their learning. P2's husband "said that because [she didn't] speak English he [would] divorce [her]." Still, two participants saw mobile translation tools as language learning tools, helping expand their vocabulary and using the text-to-speech affordances as cues for correct pronunciation:

"I could not understand English; when they assigned me to different chores I would use Google Translate to find out what it is — to get me familiarized with the vocabulary." —P4, Mandarin Chinese

"My brother, used to use the Google Translate tool to study. He used to say the words and repeat them. He would listen to the pronunciation." —P6, Spanish

### DISCUSSION

In both studies with migrant populations, lack of language proficiency had significant negative impact on individuals' lives. Although each group lived in geographically and culturally distant places, their narratives were strikingly similar. They articulated scenarios on a wide scale of complexity from simple tasks like grocery shopping to more nuanced conversations with employers and romantic relationships. When they used technology, they also needed to navigate the "gulf of execution" [31]: how to best meet their immediate language needs given the affordances of the device or app. Calling on the assistance of a bilingual friend or relative was a common strategy for all users, but some expressed privacy concerns about involving others in their affairs. For literate users, text messaging offered another asynchronous mode. Smartphone users mostly typed into the translate app and used the text and/or audio output.

Both migrant studies highlighted several significant limitations of translation technology: (1) translation errors; (2) lack of accent and dialect support as a result of deficiencies in speech model training data; (3) models lack context specific information and thus have limitations catering to vocabularies of specific contexts, e.g., occupation related vocabulary; (4) navigating device mediated conversations in high stress situations can be problematic; (5) many participants had low writing and reading literacy in their native language, indicating the need for audio and visual affordances to make language accessible.

The key functionality missing in translation apps is deep, reliable support for complex, multi-turn conversational translation. Popular translation apps including Google Translate and Microsoft Translator offer "conversation modes" for dyadic spoken conversations. The burden of configuring and using these modes falls squarely on the user. Users must still select languages, pair devices, manually initiate turns, and develop strategies for overcoming errors compounded by speech recognition and speech. The challenges of fluent speech translation are not new [40, 51] and many research directions actively pursue machine-learned model improvements. However, we propose that translation apps take a holistic view. Instead of treating speech and translation as modeling problems, consider how understanding the wide variety of translated communication can jointly inform interface and model design.

### DESIGN IMPLICATIONS

**Literacy.** In both studies, speaking proficiency was strong for all participants in their native language, but many participants self-rated their reading and writing proficiency as inadequate. This highlights the need for translation technologies to adapt to individual literacy levels and build visual and audio affordances to make language accessible. Karusala et al. [25] note that the need for audio interfaces is pervasive across applications for low-literacy populations.

Even the iconography is important. Many apps use an icon depicting the Roman letter **A** and the Chinese character 文 (*wén*, language). One participant was unfamiliar with Chinese characters and thought this symbol denoted a pair of people placing something on a table.

**Dialect and accent support.** Most speech translation systems use speech recognition transcripts piped into text translation models. Recent work shows a single, unified model can translate speech into text across languages [50]. However, the speech side of these models suffers on speech pronounced outside the domain of the training data. Our second study participants seemed to understand the underlying problem but blamed their own pronunciation:

> "You know everybody speaks different from the origin, Mexican speaks different from other countries, some of the accents, the app isn't able to understand." —P6, Spanish

We propose that model training explore regional variation in speech data across different languages to improve recognition. Some progress has been made in accented models [53], but dialect is more than just phonetic differences. Since the "standard" dialect of a language is usually determined by those in power, marginalized populations with the greatest need for language assistance will also encounter the worse performance with these models. Apps should support personalized pronunciation models: a user need only correct the recognition once to receive future benefit.

Conversely, awareness of language formality, particularly with spoken versus written language is important. Participants noted that speech translated into Tamil sounded unusual, not because it didn't make sense, but the tone was that of formal written text. UI affordances for choosing register and tone will provide the speaker with additional control.

**Design for touch-free conversation.** We noted that starting and maintaining a multi-turn conversation requires much effort on the user's part. Many results from the dialog systems community have yet to transfer into translation apps, although some commercial desktop systems have dialog managers [42]. Pressing buttons and passing devices across users detracts from the conversation flow, adds latency and obscures the side channel transmitted during face-to-face communication. We propose that translation experiences should aim to be as touch free as possible. There is inherent trust placed in an eyes-free interface, so the experience should degrade as gradually as possible. Participants used a variety of strategies to estimate the quality of the downstream translation. For literate users, viewing an accurate transcript of their own speech engenders trust in the system; one can then choose to adopt touch-free use at their own pace.

A variety of near-simultaneous speech translation devices entered the market in the late 2010s [38]. Most of these devices pair with a mobile phone and use network-connected services to perform speech recognition and translation. Most devices are built for a single person to control; the user manually initiates turn-taking through use of a physical button or capacitive surface. Although an audio-only interface addresses the aforementioned literacy issues, users will continue to experience trustworthiness issues because the transcript of the first speaker is not verifiable. Turn-taking will continue to be less natural than a system that understands human dialogue progression.

To our knowledge, there are no conversational translation systems where the system is a fluent party in the conversation. Some systems situate speech-to-speech translation in humanoid robotic form [43] and others provide structured prompts to the original speaker to disambiguate speech [36], but none emulate the clarifying questions of a professional interpreter. We propose that the app can employ a variety of strategies to improve dyadic conversation by exploiting awareness of model quality and semantics.

**Consider context.** Commerce, healthcare, and employment scenarios were common in our findings across populations. Users employ different vocabularies in these settings (e.g. [52]), so allowing the user to explicitly or implicitly bias the model based on location should improve quality. Even though some apps have favoriting features, these are global and not context-sensitive. For low-literacy users, pictographic representations could be useful.

## CONCLUSION

Cross-language communication is fundamental for many populations. We presented the results of three studies. The first, with travelers from the United States, showed relatively low frequencies of cross-language interaction relative to perceived importance. However, having a cross-language conversation with another person was most frequent and also seen as most important. The second and third studies identified scenarios encountered by migrants on a regular basis. Lack of second language skill resulted in serious consequences on a long-term basis. Individuals use translation apps in a variety of ways with varying degrees of success in scenarios from language learning to translating long-form conversations. Finally, we identified gaps in performance and proposed design implications for better matching user needs with translation app affordances. Instead of designing the app around the models, designing these apps with the user needs in mind first will provide more fluid, successful experiences for all.

## ACKNOWLEDGMENTS
We thank all participants, interpreters, and immigrant rights and community allies who supported this research. We thank Courtney Heldreth, Kristen Olson, and Nithya Sambasivan for reviewing an early manuscript.